\documentclass[pre,twocolumn,showpacs,preprintnumbers,amsmath,amssymb]{revtex4}

\usepackage{graphicx}

\begin{document}

\title{Flow and clogging in a silo with an obstacle above the orifice}

\author{Celia Lozano}

\author{Alvaro Janda}

\author{Angel Garcimart\'{\i}n}

\author{Diego Maza}

\author{Iker Zuriguel}\email{iker@unav.es}

\affiliation{Departamento de F\'{\i}sica, Facultad de Ciencias,
Universidad de Navarra, 31080 Pamplona, Spain.}

\date{\today}

\begin{abstract}
In a recent paper [Zuriguel et al., Phys. Rev. Lett. \textbf{107},
278001 (2011)] it has been shown that the presence of an obstacle
above the outlet can significatively reduce the clogging
probability of granular matter pouring from a silo. The amount of
this reduction strongly depends on the obstacle position. In this
work, we present new measurements to analyze different outlet
sizes, extending foregoing results and revealing that the effect
of the obstacle is enhanced as the outlet size is increased. In
addition, the effect of the obstacle position on the flow rate
properties and in the geometrical features of arches is studied.
These results reinforce previous evidence of the pressure
reduction induced by the obstacle. In addition, it is shown how
the mean avalanche size and the average flow rate are not
necessarily linked. On the other hand, a close relationship is
suggested between the mean avalanche size and the flow rate
fluctuations.
\end{abstract}


\maketitle

\section{Introduction}
\label{sect:intro}

When a group of particles flow through a narrow bottleneck --an
opening not much larger than the particle size-- the dissipative
interactions among the particles may lead to the spontaneous
development of clogs. These kind of jams can be observed in grains
discharging from a silo \cite{to1}, people escaping from a room
\cite{Helbing} or vehicle traffic in a highway. Although the
nature of the particles is wildly different in each one of these
instances, they share interesting resemblances which suggest that
a general theory could describe at least the most important
features of the phenomenon \cite{Helbingexponential}. One of the
most striking similarities among people escaping from a room and
beads passing through a narrowing is that the distribution of
avalanche sizes (or bursts) displays an exponential behavior.

Generally speaking, the avalanche size is defined as the number of
particles passing through the bottleneck between two consecutive
clogs. For the case of inert beads, the avalanche size is easily
determined, as the flow is halted permanently and for good by
arches formed just above the exit orifice \cite{iker1,to2,janda1}.
The flow can only be restored by breaking such arches with an
external energy input. At that moment, a new avalanche begins.
Therefore, the avalanche size is defined as the number of
particles that get out of the silo from the instant when such an
energy input is applied, till the moment when an arch blocks the
orifice and ends the outpouring. For the case of people, or other
live beings, the definition of avalanche is more problematic due
to the fact that the clogs, left to their own, last just for a
short time. Hence, the only way to define a clog is to choose a
certain time lapse during which no individual has come out from
the enclosure \cite{Saloma}. Once the clogs are thus specified,
the avalanche sizes are just measured as the number of individuals
that get out between two consecutive clogs. Obviously, in this
case the avalanche sizes depend on the time lapse chosen to
determine whether clogging has occurred, whereas in the case of
inert particles this arbitrariness is absent. This difference in
the nature of the clogs may be at the hearth of a remarkable fact:
while the exponential behavior is general in the case of granular
flows \cite{iker1,to2,janda1}, it is only observed for very small
door sizes in crowd dynamics \cite{Helbing,Saloma}.

The exponential distribution of avalanche sizes for inert beads
was explained in \cite{iker1} in terms of the probability $p$ that
a particle passes through the outlet without getting stuck.
Assuming that this probability remains constant during the whole
avalanche, the distribution of avalanche sizes was written as
\begin{equation}
\label{eq1} n(s)=p^s (1-p)
\end{equation}
where $1-p$ is, of course, the probability that a particle blocks
the orifice. The mean avalanche size $\langle s \rangle$, i.e. the
first moment of the distribution described by Eq. \ref{eq1}, can
be written as:
\begin{equation}
\label{eq2} \langle s \rangle =p/(1-p)
\end{equation}
As expected, if the probability of clogging $1-p$ increases, then
the mean avalanche size decreases.

In both cases --people escaping from a room and particles
outpouring from a silo-- the enlargement of the outlet leads to an
increase in the size of bursts (or avalanches). A lingering
question is whether there exists an outlet size above which
clogging is impossible. For the case of a silo filled with grains,
debate still goes \cite{to2,janda1}. Nevertheless, outlet sizes
larger than $7$ or $8$ times the particle size generate immense
avalanches, so in practical terms two regimes can be
distinguished: one of \emph{clogging} for small outlet sizes, and
another one of \emph{no clogging} for big outlet sizes.

Another feature of clogging that is shared by people in a crowd
and particles in a silo is the role of pressure. In humans, an
increase of pressure caused by panic seems to be a key ingredient
for the appearance of clogging \cite{Helbing}. Parisi and Corso
\cite{Parisi} have shown that the evacuation time for a room as a
function of the pedestrians' desired velocity (which is a variable
considered in the model) presents a minimum for intermediate
velocities. A very high desired velocity leads to an increase in
the evacuation time due to clogs. In granular matter, a low
pressure (which can be attained in shallow layers of grains) is
found to effectively prevent clogging \cite{prlinsert}. Low
pressure is also behind an ingenious idea sometimes used when
trying to improve the flow of pedestrians through a bottleneck:
the placement of an obstacle \emph{before} the exit
\cite{Kirchner,prlinsert,Parisi,Yanagisawa,Frank,Shiwakoti}. The
size of the column --the obstacle-- is typically of the order of a
pedestrian and the position varies from one study to another,
although it is generally close to the exit (at most at a distance
of 2 or 3 pedestrian sizes). In the case of silos, the placement
of an obstacle above the outlet has also been used, but its
relationship with clogging remains scantily explored. Instead, the
placement of obstacles is usually aimed to improve the flow and to
the reduction of undesirable wall stresses. Indeed, most of the
studies about silos and obstacles are performed with such a large
outlet size that clogging is practically impossible
\cite{Tuzun,Zelinski,Yang_insert_silos,Marroquin}.

In a previous work we reported experimental evidence of the silo
clogging reduction induced by the presence of an obstacle
\cite{prlinsert}. It was shown that the clogging probability could
be reduced by almost two orders of magnitude if the obstacle
position is properly selected. This important effect was
attributed to a pressure reduction in the outlet neighborhood.
Notably, the presumed pressure variation induced by the obstacle
affects very little the flow rate. This result is in agreement
with previous works, where it was proved that the pressure in a
silo does not have a significant influence on the flow rate
\cite{Pacheco,Aguirre_belt}. In this manuscript, we extend the
range of previous results by analyzing the effect of the obstacle
position for different sizes of both the outlet and the obstacle.
Furthermore, we investigate the effect of the obstacle in the
features of the flow rate and its fluctuations. Finally, the
comparison between the arches formed in silos with and without an
obstacle reveals meaningful geometrical differences which seem to
be related to the pressure variation. The manuscript is structured
as follows. First, we will explain the experimental setup and
procedure. Then, we will show the generality of the effect caused
by the obstacle regardless of outlet size. In the next two
sections we will analyze the flow rate properties, in connection
with clogging. Finally, we will relate the pressure reduction
induced by the obstacle with the arch shape and we will draw some
conclusions.

\section{Experimental procedure.}
\label{sect:exp}

\begin{figure}
{\includegraphics[width=0.85\columnwidth]{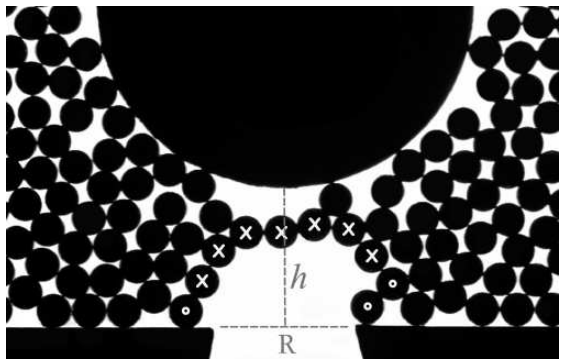}}
\caption{\label{fig:exp} Photograph of an arch formed above the
outlet. The dashed horizontal segment marks the line where the crossing of each bead is scored to compute passage times.
$R$ is the length (size) of the outlet and $h$ the distance from
the bottom of the obstacle to the outlet. Particles forming the
arch are indicated by crosses and particles forming the base of
the arch are indicated by circles.}
\end{figure}

The experimental setup consists of a two dimensional rectangular
silo 800 mm high and 200 mm wide. It is made of two glass sheets
separated by two flat metal strips $1.1\;mm$ thick and 800 mm
high. These metal bands are the side walls, so the distance
between them (200 mm) fixes the width of the silo. The silo bottom
is flat and formed by two facing metal flanges, so that their
edges define the outlet size $R$, which can be varied at will
(Fig. \ref{fig:exp}). Above the bottom, a disk of diameter $D_I$
is placed vertically above the outlet center. In most of the
experiments $D_I$ is set at $10\;mm$, although in chapter
\ref{sect:obstaclesize} this value has been modified in order to
study its effect in the discharge phenomenon. The distance $h$
from the bottom of the obstacle to the outlet (see Fig.
\ref{fig:exp}) can be varied and is carefully measured with an
accuracy of $0.05\;mm$. As in a previous work \cite{prlinsert},
the case of a silo without obstacle will be referred to as
$h\rightarrow\infty$.

The silo is filled with a sample of monodisperse stainless steel
beads with a diameter of $1.00 \pm 0.01 \;mm$. Hence, the grains
are disposed between the two glass sheets conforming a monolayer.
The silo filling is performed by pouring the grains along its
whole width through a hopper at the top. After the silo filling,
grains start to flow through the outlet until an arch blocks it.
The particles are collected in a cardboard box placed on top of a
balance. As the weight of one particle is known, the size of the
avalanche $s$ --the number of particles fallen between two
consecutive clogs-- is easily calculated. Then, a picture of the
region above the orifice is taken with a standard video camera and
further analyzed in order to detect the position of every particle
in the image. From these, the particles forming the blocking arch
are obtained as explained in \cite{Garcimartin_arches}. The
experiment is resumed by blowing a jet of compressed air aimed at
the orifice that starts a new avalanche. The experimental setup is
automated and controlled by a computer. This allows us to register
a large number of avalanches (between $800$ and $3000$) and the
corresponding arches at each run. Let us note that the silo is
refilled whenever the level of grains falls below a preset lower
limit of around $300\;mm$ ($1.5$ times the width of the silo). The
reason for this is to avoid pressure variations at the bottom due
to the amount of grains in the silo; recall that the pressure at
the base of a silo saturates and is therefore independent on the
filling level as long as the height of the granular layer exceeds
a certain level \cite{Duran}.

Additionally, for each experimental condition, a number of movies
were recorded of the region above the outlet with a high speed
camera at $1500$ frames per second during a total time lapse of
$40$ seconds. These recordings allowed to accurately measure the
moment at which each particle crosses the outlet (with a precision
better than $1$ ms). From this, the flow rate $q$ (in number of
particles per second) is calculated within time intervals of $30$
ms. It should be noted that the flow rate measurements were always
performed well inside the avalanches. In particular, we always
wait $3$ seconds after the beginning of the avalanche and we stop
the measurement at least $1$ second before the end of the
avalanche. In this way we intend to avoid any possible influence
of a transient regime at the beginning or the end of the
avalanche. We remark that, as reported before
\cite{Janda_fluctus}, it is possible to find values of $q=0$
within the avalanche. Such events were proposed to correspond to
unstable clogs: arches that interrupt the flow of the particles
for a short time but were not strong enough to durably clog the
silo. From all the values of $q$ obtained (typically $1300$
measurements for each experimental conditions) we calculated the
average flow rate $\langle q \rangle$ as well as its coefficient
of variation.

\section{Clogging reduction. Dependence on the outlet size.}
\label{sect:cloggingraduction}

\begin{figure}
{\includegraphics[width=1\columnwidth]{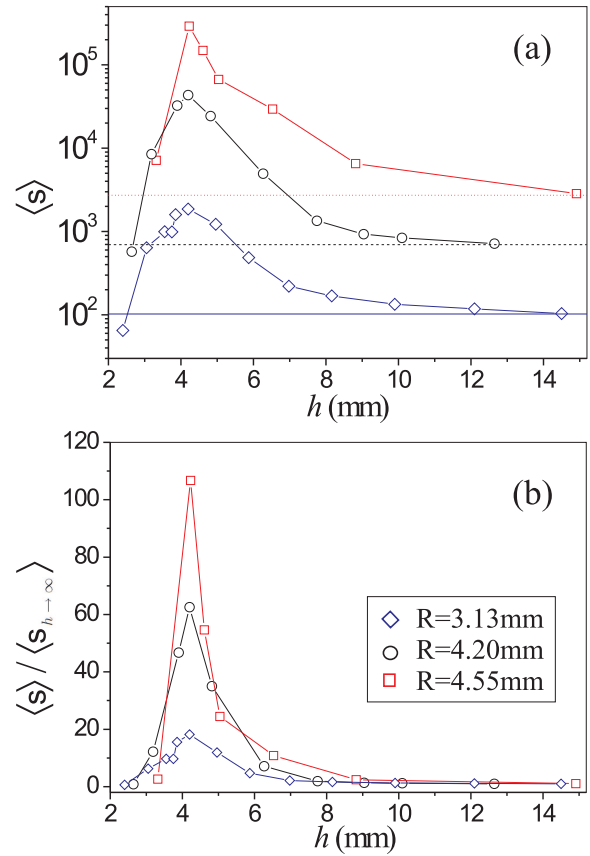}}
\caption{\label{fig:meanavas} (Color online) a) Mean avalanche
size versus $h$ for $R=3.13$ mm ($\diamond$), $R=4.20$ mm
($\circ$) and $R=4.55$ mm ($\square$). Horizontal lines indicate
the values of $\langle s \rangle$ when $h\rightarrow\infty$. b)
Same results than in (a), but dividing the mean avalanche size by
the mean avalanche size at $h\rightarrow\infty$ corresponding to
each value of $R$.}
\end{figure}

In a previous work, it was shown that the avalanche size
distribution displays an exponential decay for all the obstacle
positions \cite{prlinsert}. Hence, the mean avalanche size
$\langle s \rangle$ can be defined and used as a characteristic
parameter of the distribution. Let us now focus on the behavior of
$\langle s \rangle$ as a function of the obstacle height above the
outlet ($h$) for different outlet sizes. In Fig.
\ref{fig:meanavas} (a) we plot the experimental values of $\langle
s \rangle$ versus $h$ for three different outlet sizes ($R=3.13$,
$4.20$, and $4.55$ mm). For these values of $R$, the mean
avalanche size if the obstacle is absent ($\langle
s_{h\rightarrow\infty} \rangle$) extends over a wide range: from
$\langle s_{h\rightarrow\infty} \rangle=100$ to $\langle
s_{h\rightarrow\infty} \rangle=3000$ particles. These values are
represented by dashed horizontal lines in Fig. \ref{fig:meanavas}
(a). This plot manifests that the mean avalanche size as a
function of $h$ displays the same trend independently of the
outlet size. In all the cases, for positions of the obstacle above
$10$ mm or so, the value of $\langle s \rangle$ is very similar to
the value obtained without obstacle. This indicates that for such
values of $h$ the obstacle effect nearly vanishes. When the
obstacle approaches the orifice from $h\rightarrow\infty$,
$\langle s \rangle$ increases, displaying a maximum for
$h\approx4$ mm. For smaller values of $h$ the mean avalanche size
decreases, because arches begin to build up between the obstacle
and the bottom (before that, arches just span over the exit
orifice without touching the obstacle) \cite{prlinsert}.

Despite the similar behavior displayed in Fig. \ref{fig:meanavas}
(a) for different outlet sizes, it is evident that the consequence
of the peak is enhanced as $R$ increases. This result is more
obvious if we plot $\langle s \rangle$ divided by the values of
$\langle  s_{h\rightarrow\infty} \rangle$ corresponding to each
$R$ (Fig. \ref{fig:meanavas}b). This behavior makes sense, because
as $R$ increases and approaches values of `no clogging', arches
are composed of more particles and are weaker \cite{Mankoc,
Lozanosubmitted}. Hence, a similar decrease of pressure can result
in a stronger reduction of the clogging probability (or
enhancement of the mean avalanche size) as $R$ increases. Another
interesting fact observed in Fig. \ref{fig:meanavas}b is that the
position of the peak does not depend significatively on the outlet
size. This result can be understood if we assume that the decay of
$\langle s \rangle$ for small values of $h$ is due to clogs
developed between the orifice and the obstacle. Thus, in this
region, the parameter that governs clogging is not $R$ but the
distance between the bottom of the obstacle and the bottom of the
silo. In any case, the mechanisms causing clogging for such small
values of $h$ --which should be closely related to clogging in
inclined orifices \cite{Sheldon,ThomasandDurian}-- are not the
focus of this paper. In brief, if we just consider the domain
where arches are not formed between the outlet and the obstacle
($h>4$), it can be stated that the bigger the outlet, the stronger
the effect of the obstacle in the clogging reduction.

\section{Flow rate}
\label{sect:flowratefluctus}

\begin{figure}
{\includegraphics[width=1\columnwidth]{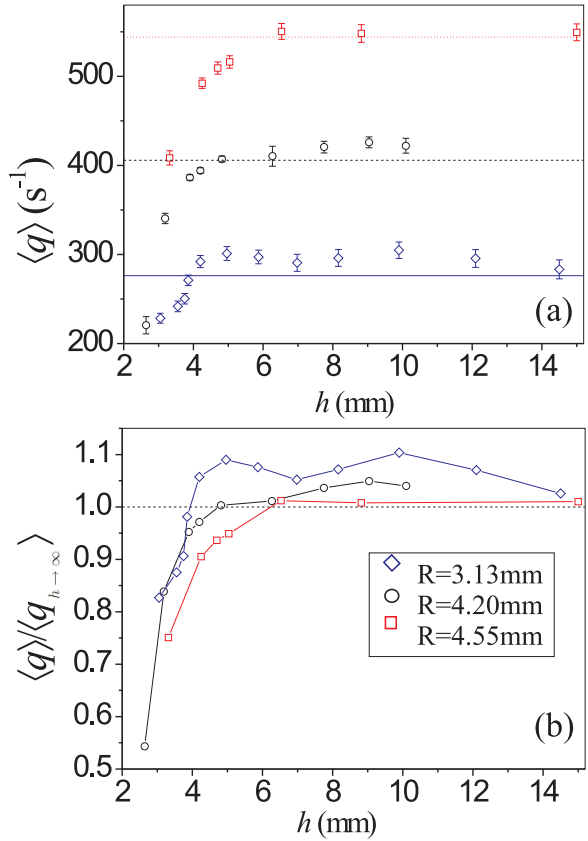}}
\caption{\label{fig:flowrate}(Color online) a) Average flow rate
versus $h$ for $R=3.13$ mm ($\diamond$), $R=4.20$ mm ($\circ$) and
$R=4.55$ mm ($\square$). Error bars are confidence intervals at
$95\%$. Horizontal lines indicate the values of $\langle q
\rangle$ when $h\rightarrow\infty$. b) Same results than in a) but
dividing the average flow rate by the average flow rate at
$h\rightarrow\infty$ corresponding to each value of $R$.}
\end{figure}

In this section we will analyze the effect of the obstacle on the
flow rate (calculated as explained in section \ref{sect:exp}). We
first show in Fig. \ref{fig:flowrate}a the results of the average
flow rate $\langle q \rangle$ obtained for different obstacle
positions $h$. Clearly, the behavior is similar for the three
outlet sizes ($R$) studied in this work. When $h$ is too small,
the flow rate is smaller than the one obtained without obstacle
(which is marked with dashed horizontal lines in Fig.
\ref{fig:flowrate}a). This is due to the short distance between
the outlet and the obstacle, that strongly affects the flow rate.
If the obstacle position is moved upwards, far from the orifice,
the flow rate can be increased, reaching values up to $10\%$
higher than in the silo without obstacle. This result is in
agreement with recent results obtained for large outlet sizes
where clogging did not appear \cite{Yang_insert_silos,Marroquin}.
The measurements displayed in Fig. \ref{fig:flowrate}a also
indicate that the flow rate enhancement is more conspicuous for
the smallest outlet size. In addition, it seems that the
transition point from flow rate reduction (at small values of $h$)
to flow rate enhancement (at high values of $h$) moves towards
higher values of $h$ as $R$ is increased. This is more clearly
seen if we represent the results of the flow rate rescaled by the
flow rate obtained without obstacle corresponding to each outlet
size, \emph{i. e.} $\langle q \rangle / \langle
q_{h\rightarrow\infty} \rangle$ (see Fig. \ref{fig:flowrate} b).

We remark that no obvious relationship can be perceived between
the average flow rate and the mean avalanche size measurements.
Indeed, the effect of the obstacle on the avalanche size is more
prominent as $R$ is increased, whereas the effect on the flow rate
is stronger for small $R$. At the same time, the obstacle
positions at which the maximum flow rate is obtained do not
coincide with the positions at which the avalanche size is
maximized. All these facts suggest the different nature of two
processes: the flow of particles through the outlet, and the
clogging due to arch formation \cite{Jandasubmitted}.

\begin{figure}
{\includegraphics[width=1\columnwidth]{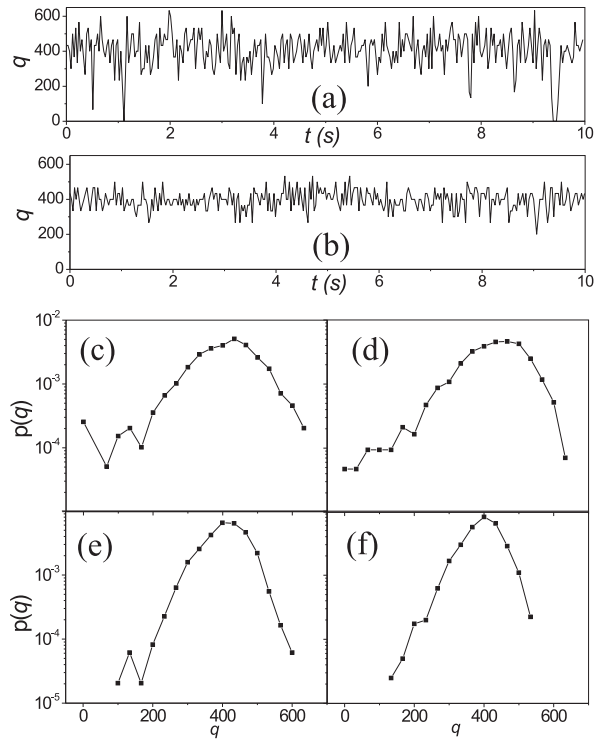}}
\caption{\label{fig:flowratefluctus} Time series of the flow rate
q (beads per unit time as measured at nonoverlapping time windows
of 30 ms) for an orifice size $R=4.2$ and an obstacle of $10$ mm
diameter placed at a) $h\rightarrow\infty$ and b) $h=4.20$ mm. In
figures c-e the normalized histograms of the flow rate values are
presented for different obstacle positions, namely: c)
$h\rightarrow\infty$, d) $h=9.04$ mm, e) $h=4.82$ mm and f)
$h=4.20$ mm.}
\end{figure}

On the other hand, it has been recently proposed that some
connection does exist between the mean avalanche size and the
fluctuations of the flow rate \cite{Janda_fluctus}. Indeed, it was
shown that in a 2D silo the values of $q$ display a Gaussian
distribution if the outlet size is large. On the contrary, as $R$
was reduced and the region of clogging approached, there was an
increase on the number of events with $q\approx0$, so that the
distribution was no longer Gaussian. Those events were attributed
to the existence of partial clogs that were not strong enough to
permanently halt the flow. In Fig. \ref{fig:flowratefluctus}a we
present the time series of $q$ obtained for a silo with orifice
size $R=4.2$ without obstacle. In agreement with
\cite{Janda_fluctus}, the trace displays downward spikes in which
the flow goes to zero. These events are consequence of temporal
(not definitive) interruption of the flow. If the same results are
presented for the case of a silo with an obstacle of 10 mm
diameter placed at $h=4.2$ (Fig. \ref{fig:flowratefluctus} b) it
becomes clear that the downward spikes disappear. Recall that
$h=4.2$ is the obstacle position for which the avalanche size is
maximum and hence, the clogging probability is minimum. In
addition, from Fig. \ref{fig:flowratefluctus} b, it seems that
placing an obstacle also minimizes the upward fluctuations making
the flow rate more homogeneous while its average value is
practically unaltered.

In Figs. \ref{fig:flowratefluctus}c-f we present the normalized
distributions of $q$ for four different obstacle positions. Fig.
\ref{fig:flowratefluctus}a shows the distribution for $q$ without
obstacle, where the existence of a large number of events with
$q\approx0$ is rather obvious. As the position of the obstacle is
moved toward the outlet, the number of events corresponding to
$q\approx0$ is reduced, to the point of being almost absent when
$h=4.2$ (Fig. \ref{fig:flowratefluctus}d). In addition, decreasing
$h$, the histograms become narrower implying that the placement of
the obstacle helps to avoid fluctuations of the flow rate. The
results obtained for the distributions of $q$ for the two other
outlet sizes (not shown) display the same behavior.

\begin{figure}
{\includegraphics[width=0.9\columnwidth]{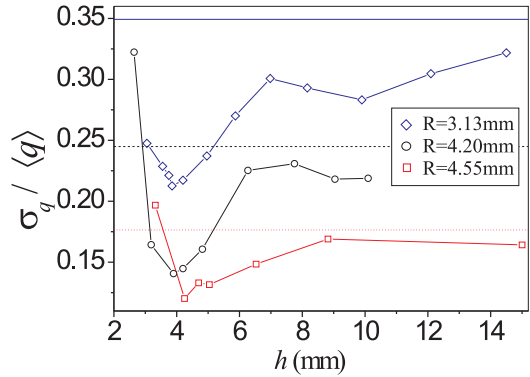}}
\caption{\label{fig:sigmaflowrate} (Color online) Coefficient of variation, \emph{i.e.} standard deviation of the values obtained for the flow rate $q$ rescaled by $\langle q \rangle$, versus the obstacle position for $R=3.13$ mm ($\diamond$), $R=4.20$ mm ($\circ$) and $R=4.55$ mm ($\square$). The horizontal lines indicate the values of $\sigma_q/\langle q \rangle$ for $h\rightarrow\infty$ for each value of $R$.}
\end{figure}

The effect of the obstacle position in the flow rate fluctuations
is quantified by means of the standard deviation of the distribution divided by the average, \emph{i. e.} the coefficient of variation $CV = \sigma_q/\langle q \rangle$. The results obtained for different obstacle positions and the different values of $R$ are presented in Fig. \ref{fig:sigmaflowrate}. Interestingly, the global behavior obtained can be easily related to the one reported in Fig.
\ref{fig:meanavas} a for the mean avalanche size. For all the
outlet sizes studied, high values of $h$ display values of $CV$
close to the ones seen in the absence of an obstacle. As $h$ is
reduced, the fluctuations decrease just as the probability of
clogging does (which implies an increase in the mean avalanche
size). In all the cases, the minimum of $CV$ is reached at values
of $h$ similar to those for which the maximum avalanche size is
obtained. These results exhibit the close relationship between clogging reduction and the fluctuations of the flow rate: reducing the clogging probability is always associated with a reduction in the fluctuations of the flow rate.

Let us finally stress that by placing an obstacle above the
outlet, the flow rate fluctuations can be reduced without an
associated increase of the outlet size and average flow rate. This
implies a clear advantage with respect to the work of Janda et al.
\cite{Janda_fluctus}, where the reduction of fluctuations was
always associated with an increase of the flow rate.

\section{Obstacle size}
\label{sect:obstaclesize}

\begin{figure*}
{\includegraphics[width=1.9\columnwidth]{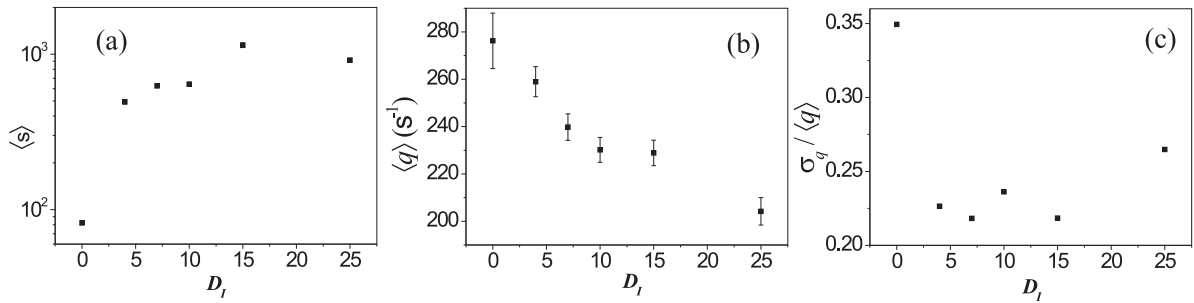}}
\caption{\label{fig:obstaclesize} (a) Mean avalanche size, (b)
average flow rate, and (c) coefficient of variation for different
obstacle diameters ($D_I$). In all the graphs $D_I=0$ represents
the values obtained for the silo without obstacle. All the
obstacles were placed at $h=3.05 \pm 0.07$ mm in a silo where the
outlet size was $R=3.13 \pm 0.05$ mm. Error bars (which are of the
size of the symbols in Fig. a) are confidence intervals at
$95\%$.}
\end{figure*}

In order to further explore the possible relationship between
clogging and flow rate fluctuations, we have performed some
experiments where the diameter of the obstacle ($D_I$) has been
varied. We choose a fixed outlet size ($R=3.13 \pm 0.05$ mm) and
obstacle position ($h=3.05 \pm 0.07$ mm). The election of $R$ is
based in practical terms as the mean avalanche size is small, with
the consequent time savings that this implies. The reason for
choosing $h=3.05 \pm 0.07$ mm is that, for such values of $h$,
both the avalanche size and the average flow rate depend on the
obstacle position (Fig. \ref{fig:meanavas} and Fig.
\ref{fig:flowrate} respectively). Therefore, we expect that the
change of obstacle size has also an effect on these variables. We
suspect that choosing a higher value of $h$ could lead to values
of the flow rate and avalanche size less dependent on the obstacle
size. In any case, this hypothesis remains to be confirmed in
future works.

In Fig. \ref{fig:obstaclesize}a the results of the mean avalanche
size are displayed for five different diameters of the obstacle
and compared with the case without obstacle (which is presented as
$D_I=0$). The first result that becomes evident is that the
obstacle, whatever its size, causes an important increase of the
mean avalanche size. The mere presence of the obstacle, even when
it is small, prevents the clogging of particles passing through a
bottleneck --at least for the outlet size and the obstacle
position used here. In addition, it seems that there is a obstacle
size ($D_I\approx15$ mm) for which the avalanche size is maximum.
This result can be understood as follows: for small obstacle
sizes, the bigger the obstacle, the stronger its effect preventing
clogging. Hence, enlarging the obstacle leads to an increase of
the avalanche size. This tendency is reversed for very big
obstacle sizes, a fact which is attributed to an enhancement of
the clogging between the obstacle and the bottom of the silo.
Intuitively, an exceedingly large obstacle would lead to a
situation where the flow would be impossible, as the angle of
repose of the material imposes a limit at which the particles are
sustained by themselves \cite{ThomasandDurian}.

We also analyzed if the flow rate depends on the obstacle size;
measurements are presented in Fig. \ref{fig:obstaclesize}b. It can
be seen that the larger the diameter of the obstacle, the smaller
the average flow rate. This behavior can be understood if we think
that increasing the obstacle size leads to a reduction of the
distance between the obstacle and the static region of grains at
the bottom corners of the silo imposed by the angle of repose.
This process is analogous to moving the obstacle towards the
orifice (decreasing $h$). Let us recall that the obstacles of
different diameter were placed at a position ($h=3.05 \pm 0.07$
mm) for which the average flow rate decreases as the obstacle
approaches the outlet (as shown in Fig. \ref{fig:flowrate}).

Finally, in Fig. \ref{fig:obstaclesize}c we show the variation
coefficient as a function of $D_I$. Interestingly, the highest
value of $\sigma_q/\langle q \rangle$ is obtained for the silo
without obstacle, where the minimum value of the mean avalanche
size is also found. In addition, the coefficient of variation also
displays the non-monotonic dependence on $D_I$ that was observed
for the mean avalanche size, supporting the suggestion of Sect.
\ref{sect:flowratefluctus} about the relationship of reduced
clogging probability and lessened flow rate fluctuations. On the
contrary, the average flow rate and the mean avalanche size (or
the clogging probability) do not seem to be related in any
recognizable way.

\section{Arch shape}
\label{sect:archshape}

From the pictures recorded in the experiment (Fig. \ref{fig:exp})
the position of the beads in the arches that block permanently the
silo can be accurately obtained. As reported in a previous paper
\cite{Garcimartin_arches}, the particles forming the base of the
arch are not considered to belong to it, in compliance with the
definition of arch given in numerical works
\cite{Pugnaloni_arches}. In practical terms, the particles in the
base are those whose centers are at the extreme positions in the
horizontal direction. In Fig. \ref{fig:exp} the centers of the
particles belonging to the arch are marked with crosses and the
particles of the base are marked with circles. Once the particles
forming the arch have been identified, a number of arch properties
can be analyzed (the size, measured in number of beads; span;
height; and aspect ratio). The span is defined as the distance,
projected on the horizontal direction, between the two outermost
particles of the arch. Accordingly, the height is defined as the
distance between the vertical coordinates of the centers
corresponding to the highest and the lowest beads of the arch.
Finally, the aspect ratio ($A$) is calculated as the quotient
between half the span and the height of the arch. All these
parameters will allow to explore if the presence of an obstacle
induces any kind of difference in the geometrical properties of
the arches. In this work, we will compare the arch features in a
silo with an orifice of $R=4.2$ mm in two different situations:
without obstacle and with an obstacle at $h=4.2$ mm. We note that
for this value of $R$, the obstacle placed at $h=4.2$ mm yields
the maximum avalanche size observed (Fig. \ref{fig:meanavas} a).

\begin{figure}
{\includegraphics[width=1\columnwidth]{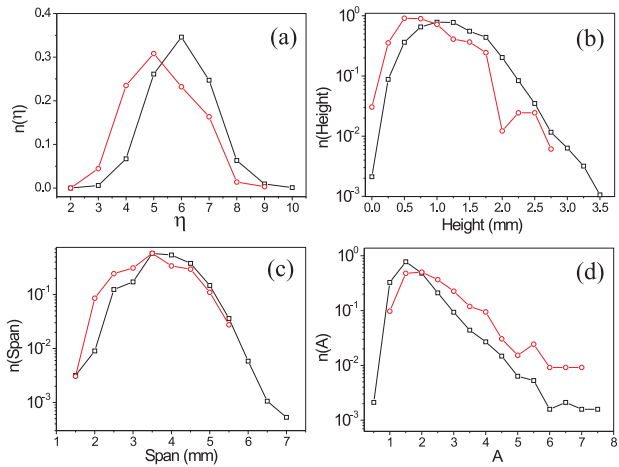}}
\caption{\label{fig:shapearches} (Color online) (a) Normalized
histogram of the size of the arches measured in number of beads
($\eta$). (b) Normalized histogram of the arch height (note the semilogarithmic scale). (c) Normalized histogram of the arch span
(in semilogarithmic scale). (d) Normalized histogram (in
semilogarithmic scale) of the aspect ratios (A) of the arches,
defined as half the span divided by the height. In all the graphs
measurements have been carried out in a silo with an outlet size $R=4.2$ mm, without an obstacle ($\square$) and with a $10$ mm diameter obstacle placed at $h=4.2$ mm ($\circ$).}
\end{figure}

For these two cases, namely, no obstacle and an obstacle placed at
$h=4.2$ mm, let us compare the arch size, measured in number of
beads ($\eta$). In Fig. \ref{fig:shapearches}a we present the two
arch size distributions. Clearly, smaller arches (with fewer
particles) are formed when there is an obstacle. The arch height
distribution (Fig. \ref{fig:shapearches}b) and the arch span
distribution (Fig. \ref{fig:shapearches}c) confirm this result: in
the presence of an obstacle the arches are shorter and narrower.
The aspect ratio $A$, plotted in Fig. \ref{fig:shapearches}d,
shows how arches are flatter (higher $A$) when there is an
obstacle. In addition, it seems that both distributions present an
exponential decay in the aspect ratio probability.

The aspect ratio of an arch is a especially important variable, as
it is used as a key ingredient in one of the few models linking
the silo clogging probability to the arch properties
\cite{to1,To_semicircle}. In this model the arch is proposed to be
semicircular, i.e. an aspect ratio equal to one. In an
experimental work performed afterwards, the validity of this
assumption was demonstrated for large arches (broader than the
outlet by more than one bead diameter) \cite{Garcimartin_arches}.
In addition, the aspect ratio of the arches is appealing because
it may reflect interesting features related with the loads
sustained by them. Thus, an aspect ratio smaller than one (pointed
arch) would indicate that the arch is optimized to sustain a
vertical pressure. On the contrary, an arch optimized to sustain a
horizontal load would be flatter, exhibiting an aspect ratio
larger than one. A semicircular arch (aspect ratio one) is the
preferred shape to optimize an isotropic pressure
\cite{Osserman,Ochsendorf}. Based on these assumptions, the
results presented in Fig. \ref{fig:shapearches}d are consistent
with the fact that an obstacle above the outlet screens the
pressure of the particles in the silo. If so, the flatter arches
(higher aspect ratio) obtained when an obstacle is present could
be the consequence of the load reduction in the vertical direction
caused by the obstacle.

\begin{figure}
{\includegraphics[width=1\columnwidth]{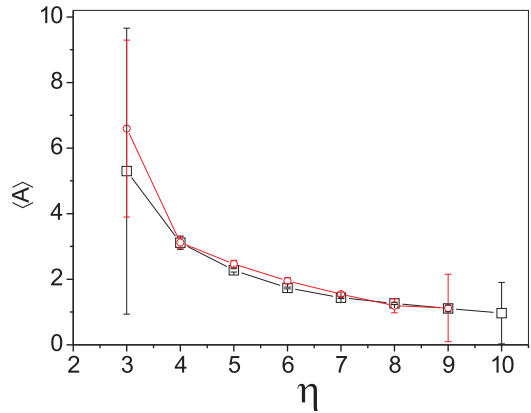}}
\caption{\label{fig:Avsnumberparticles} (Color online) Average
aspect ratio for arches as a function of their number of particles ($\eta$). As in Fig. \ref{fig:shapearches}, the measurements have been carried out in a silo with an outlet size $R=4.2$ mm, without an obstacle ($\square$) and with a $10$ mm diameter obstacle placed at $h=4.2$ mm ($\circ$). The error bars are the $95\%$ confidence intervals.}
\end{figure}

Once we have seen that the arches developed in the presence of an
obstacle are smaller (in number of particles, height and span) and
flatter, we can consider if those are independent effects or else
if one effect is caused by the other. Indeed, in a previous work
\cite{Garcimartin_arches} it was suggested that all these
variables are strongly related. In Fig.
\ref{fig:Avsnumberparticles} we present the values of the aspect
ratios averaged over all the arches formed by a given number of
particles. The results obtained for silos without an obstacle and
with an obstacle at $h=4.2$ mm are almost identical and confirm
that the aspect ratio of the arches is one when they are
sufficiently large. The only difference between both situations is
that arches in the silo without obstacle are formed by a higher
number of particles --reaching maximum values of 10 instead of 9--
something that was expected from Fig.\ref{fig:shapearches}. The
fact that the obtained results in both situations are so far
indistinguishable, seems to indicate that the arches of a given
number of particles have the same geometrical properties
regardless of the presence of an obstacle. The only effect of
placing an obstacle is to prevent the formation of pointed arches
(aspect ratio smaller than one) which, in general, consist of more
particles. It cannot be conclusively asserted whether the aspect
ratio is responsible for this behavior --the cause could instead
be ascribed to other variable--, but the above mentioned
relationship between the aspect ratio and the load distribution
makes it a likely candidate.

\section{Conclusions}
\label{sect:conclusions}

We have presented a detailed analysis of the effect that placing
an obstacle above the outlet of a silo has on the flow and
clogging processes. The measurement of the dependence of the
avalanche size on the obstacle position, and this for different
outlet sizes, reveals that the clogging reduction caused by the
obstacle is enhanced as the outlet size is increased, thereby
approaching the region of `no clogging'. This phenomenon can be
understood in terms of the robustness of the arches. If we assume
that arches become weaker as their size increases, it seems
plausible that a similar pressure reduction has a stronger effect
in greater arches than in smaller ones. Although it is still not
clear if there is a critical outlet size above which clogging is
forbidden \cite{to2,janda1}, if we assume that this boundary does
exist in practical terms, then it makes sense that small
perturbations have more consequence as the critical point is
approached.

We have also presented results of the flow rate properties for
different outlet sizes and obstacle positions. Concerning the
average flow rate, it is shown that the obstacle can cause an
increase up to around $10\%$. This increase of the flow rate is
shown to be relatively more pronounced as the outlet size is
reduced. In other words, the placement of the obstacle has a
stronger effect in the flow rate for small outlets. This trend is
opposite to the one observed for the avalanche size (or the
clogging probability). Indeed, it seems that there is not any
relationship between the average flow rate and the probability of
clogging. On the contrary, we report a clear and strong
relationship between the clogging reduction (increasing of
avalanche size) and a decrease of the coefficient of variation in
the flow rate fluctuations. This could be of practical interest as
it suggests that the measurement of this coefficient of variation
would be enough to estimate the mean avalanche size. The
relationship among these two parameters may be signaling that the
number of stable and unstable clogs is correlated. Increasing the
number of stable clogs leads to a decrease of the avalanche size,
while increasing the number of unstable clogs provokes an
enhancement of the fluctuations. This behavior was already
observed in previous works \cite{Janda_fluctus} by varying the
outlet size in a silo without an obstacle: by increasing the
outlet size, the probability of clogging was reduced, as well as
the fluctuations, but the price to pay was a corresponding
increase of the average flow rate. The interesting fact observed
when placing an obstacle is that fluctuations and average flow
rate are not necessarily linked. The use of different obstacle
sizes has allowed us to check that this effect is robust.

Finally, we have performed an exhaustive comparison of the shape
of the arches developed in a silo in the two situations where the
pressure difference is apparently more significant: the silo
without an obstacle and the silo with an obstacle placed at
$h=4.2$ mm. The results reveal that smaller and flatter arches are
obtained in the silo with an obstacle --the case in which the
pressure at the outlet is reduced. This can be qualitatively
understood if we recall that pointed arch shapes optimize vertical
loads (which seem to be more important in the absence of an
obstacle) and flatter arches are the optimal response to
horizontal loads (which seem to be more important in the silo with
obstacle).

In summary, the experimental results presented in this work show
that the placement of an obstacle has a robust effect in the silo
clogging reduction. In addition, we report convincing evidences
that the mechanism by which this effect is attained is a pressure
reduction near the orifice. However, other possibilities --such as
a modification of the flow streamlines similar to the one reported
in \cite{Kolb}-- should not be discarded. Another interesting
analogy of this work can be drawn with experiments where a big
obstacle moves slowly within a granular media \cite{Albert,Reddy}.
Indeed in \cite{Reddy} it is suggested that the stress
fluctuations induced by an obstacle in a dense granular flow close
to the jamming transition, may help to overcome the flow
threshold. Further experiments and simulations would be
interesting in order to characterize the pressure exerted by the
granular media on the obstacle, as well as its fluctuations.

\section{acknowledgments}

This work has been financially supported by Projects
FIS2008-06034-C02-01 and FIS2011-26675 (Spanish Government) and by
PIUNA (Universidad de Navarra). C.L. thanks Asociaci\'on de Amigos
de la Universidad de Navarra for a scholarship.


\begin{thebibliography}{19}



\bibitem{to1}
K. To, P.-Y. Lai and H. K. Pak, Phys. Rev. Lett. \textbf{86}, 71 (2001).

\bibitem{Helbing} D. Helbing, L. Buzna, A. Johansson, T. Werner, Transportation
Science \textbf{39}, 1 (2005). 

\bibitem{Helbingexponential}  D. Helbing, A. Johansson, J. Mathiesen, M. H. Jensen, and A. Hansen, Phys. Rev. Lett. \textbf{97},
168001 (2006). 



\bibitem{iker1}
I. Zuriguel, A. Garcimart\'{\i}n, D. Maza, L. A. Pugnaloni and J.
M. Pastor, Phys. Rev. E \textbf{71}, 051303 (2005); I. Zuriguel,
L. A. Pugnaloni, A. Garcimart{\'\i}n, and D. Maza, Phys. Rev. E
\textbf{68}, 030301 (R) (2003).

\bibitem{to2} K. To, Phys. Rev. E \textbf{71}, 060301 (R) (2005).

\bibitem{janda1} A. Janda, I. Zuriguel, A. Garcimart\'{\i}n, L. A. Pugnaloni and D.
Maza, Europhys. Lett. \textbf{84}, 44002 (2008).


\bibitem{Saloma} C. Saloma, G. J. Perez, G. Tapang, M. Lim, C. Palmes-Saloma, Proc. National Acad. Sci. USA (PNAS) \textbf{100}, 11947 (2003).


\bibitem{Parisi} D.R. Parisi, and C.O. Dorso, Physica A \textbf{354}, 606 (2005).

\bibitem{prlinsert} I. Zuriguel, A. Janda, A. Garcimart\'{\i}n, C. Lozano, R. Ar\'evalo, and D.
Maza, Phys. Rev. Lett. \textbf{107}, 278001 (2011).

\bibitem{Kirchner} A. Kirchner, K. Nishinari, and A. Schadschneider,
Phys. Rev E \textbf{67}, 056122 (2003).

\bibitem{Yanagisawa} D. Yanagisawa, A. Kimura, A. Tomoeda,
R. Nishi, Y. Suma, K. Ohtsuka, and K. Nishinari, Phys. Rev. E
\textbf{80}, 036110 (2009).


\bibitem{Frank} G.A. Frank, C.O. Dorso, Physica A \textbf{390}, 2135
(2011).

\bibitem{Shiwakoti} N. Shiwakoti, M.Sarvi, G. Rose, M. Burd, Transportation Research Part B: Methodological \textbf{45}, 1433
(2011); N. Shiwakoti, M.Sarvi, G. Rose, M. Burd, Transportation
Research Record \textbf{2137}, 31 (2009).



\bibitem{Tuzun} U. T\"uz\"un and R. M. Nedderman, Chem. Eng. Sci. \textbf{40}, 325 (1985).

\bibitem{Zelinski} B. Zelinski, E. Goles and M. Markus, Physics of
Fluids \textbf{21}, 031701 (2009). 


\bibitem{Yang_insert_silos} S. C. Yang and S. S. Hsiau, Powder Technol. \textbf{120}, 244
(2001). 

\bibitem{Marroquin}  F. Alonso-Marroquin, S.I. Azeezullah and S.A. Galindo-Torres and L.M.
Olsen-Kettle, Phys. Rev. E \textbf{85}, 020301(R) (2012).



\bibitem{Aguirre_belt} M. A. Aguirre, J. G. Grande, A. Calvo, L. A. Pugnaloni, and J.-C. G\'eminard, Phys. Rev. Lett \textbf{104}, 238002
(2010); M. A. Aguirre, J. G. Grande, A. Calvo, L. A. Pugnaloni,
and J.-C. G\'eminard, Phys. Rev. E \textbf{83}, 061305 (2011).

\bibitem{Pacheco} H. Pacheco-Martinez, H.J. van Gerner, and J. C. Ruiz-Su\'arez, Phys. Rev E \textbf{77}, 021303 (2008).





\bibitem{Garcimartin_arches} A. Garcimart\'{\i}n, I. Zuriguel, L.A. Pugnaloni, and A.
Janda, Phys. Rev. E \textbf{82}, 031306 (2010).

\bibitem{Duran} J. Duran, \emph{Sables, Poudres et Grains} (Eyrolles, Paris, 1997).

\bibitem{Mankoc} C. Mankoc, A. Garcimart\'{\i}n, I. Zuriguel, D. Maza and L.
A. Pugnaloni, Phys. Rev. E \textbf{80}, 011309 (2009).

\bibitem{Lozanosubmitted} C. Lozano, G. Lumay, I. Zuriguel, R.C. Hidalgo, and A.
Garcimart\'{\i}n, Phys. Rev. Lett \textbf{109}, 068001 (2012).

\bibitem{Sheldon} H. G. Sheldon and D. J. Durian, Granular Matter \textbf{12}, 579 (2010).

\bibitem{ThomasandDurian} C. C. Thomas and D. J. Durian, arXiv:1206.7052v1
(cond-mat.soft).

\bibitem{Jandasubmitted} A. Janda, I. Zuriguel, D. Maza, Phys. Rev. Lett \textbf{108}, 248001 (2012).

\bibitem{Janda_fluctus} A. Janda, R. Harich, I. Zuriguel, D. Maza, P. Cixous, and
A. Garcimart\'{\i}n, Phys. Rev. E \textbf{79}, 031302 (2009).

\bibitem{Pugnaloni_arches} L. A. Pugnaloni and G. C. Barker, Physica
A \textbf{337}, 428 (2004).


\bibitem{To_semicircle} K. To and P.-Y. Lai, Phys. Rev. E \textbf{66}, 011308
(2002).

\bibitem{Osserman} R. Osserman, Not. Am. Math. Soc. \textbf{57}, 220
(2010).

\bibitem{Ochsendorf} J. Ochsendorf (private communication).

\bibitem{Kolb} E. Kolb, J. Cviklinski, J. Lanuza, P. Claudin, and E. Cl\'ement, Phys. Rev. E \textbf{69}, 031306 (2004).

\bibitem{Albert} R. Albert, M.A. Pfeifer, A.-L. Barab\'asi, and P. Schiffer, Phys. Rev. Lett \textbf{82}, 205 (1999).

\bibitem{Reddy} K.A. Reddy, Y. Forterre, and O. Pouliquen, Phys. Rev. Lett \textbf{106}, 108301 (2011).




\end{thebibliography}
\end{document}